\documentclass[graybox]{svmult}

\usepackage{type1cm}        % activate if the above 3 fonts are
\usepackage{makeidx}         % allows index generation
\usepackage{graphicx}        % standard LaTeX graphics tool
\usepackage{multicol}        % used for the two-column index
\usepackage[bottom]{footmisc}% places footnotes at page bottom

\usepackage{newtxtext}       % 
\usepackage{cite}
\makeindex             % used for the subject index
\begin{document}
\title*{Intelligent interactive technologies for mental health and well-being}
% Use \titlerunning{} for an abbreviated version of
\author{Mladjan Jovanovic, Aleksandar Jevremovic, Milica Pejovic-Milovancevic}
% Use \authorrunning{Name} for an abbreviated version of
\institute{Mladjan Jovanovic \at Singidunum University, Belgrade, Serbia, \email{mjovanovic@singidunum.ac.rs}
\and Aleksandar Jevremovic \at Singidunum University, Belgrade, Serbia, \email{ajevremovic@singidunum.ac.rs}
\and Milica Pejovic-Milovancevic \at Institute of Mental Health and Faculty of Medicine, University of Belgrade, Serbia, \email{milica.pejovic@imh.org.rs}}
\maketitle

\abstract*{Mental healthcare has seen numerous benefits from interactive technologies and artificial intelligence. Various interventions have successfully used intelligent technologies to automate the assessment and evaluation of psychological treatments and mental well-being and functioning. These technologies include different types of robots, video games, and conversational agents. The paper critically analyzes existing solutions with the outlooks for their future. In particular, we: i) give an overview of the technology for mental health, ii) critically analyze the technology against the proposed criteria, and iii) provide the design outlooks for these technologies.}

\abstract{\footnote{This a pre-print version of the paper published in: E. Pap (ed.), Artificial Intelligence: Theory and Applications,
Studies in Computational Intelligence 973, Springer Nature Switzerland AG 2021,
DOI: https://doi.org/10.1007/978-3-030-72711-6\_18} Mental healthcare has seen numerous benefits from interactive technologies and artificial intelligence. Various interventions have successfully used intelligent technologies to automate the assessment and evaluation of psychological treatments and mental well-being and functioning. These technologies include different types of robots, video games, and conversational agents. The paper critically analyzes existing solutions with the outlooks for their future. In particular, we: i) give an overview of the technology for mental health, ii) critically analyze the technology against the proposed criteria, and iii) provide the design outlooks for these technologies.}

\keywords{Human-AI interaction, intelligent systems, digital healthcare, mental health, mental well-being, review, survey, robotic technologies, video games, conversational agents, chatbots, affective computing.}

\section{Introduction}
\label{sec:1}

Artificial Intelligence (AI) is among the oldest disciplines of computer science that builds systems that resemble humans in learning, thinking, problem-solving, and decision making. The field received significant interest in the previous decade, mainly due to advances in automated machine learning (ML) and deep learning (DL). They learn useful patterns from a large amount of data and keep the acquired knowledge as model structures and parameters that can be further applied to make predictions by interpreting unseen data \cite{r1}. These models are either a set of elements or features that contribute when making decisions (in ML) or are organized into several layers of abstraction (such as neural networks) for general and specific interpretation tasks (in DL).

Healthcare provision and medicine are one of the most significant challenges of AI due to being the pillars for a global society and the necessity of providing higher-quality assistance to the healthcare workforce \cite{r2}. 

An emerging and expanding domain for for application of AI is mental health. Readily available and ubiquitous devices and applications enable the provision of flexible mental care - on-demand, at any time, both at healthcare facilities and at home. Societal challenges, such as the ongoing Covid-19 pandemic, can necessitate physical distancing. Consequently, prolonged social isolation brings common risks for mental health and well-being, potentially triggering conditions such as loneliness, anxiety, and depression \cite{r3, r4}. The growing demand for telemedicine as contactless, remote, and accessible healthcare services can help users stay connected, visit their doctors remotely, and self-manage their mental functioning during the global Covid-19 lockdown \cite{r3}.

As they move into the digital era, mental healthcare is seeing substantial benefits from interactive technologies and ML/DL, including treatment both at home and in hospitals. Medical decision-making is by its very nature uncertain, unknown, inconsistent, and lacking data from multi-dimensional spaces. At the same time, it must capture patients' heterogeneity to adjust healthcare decisions, prescriptions, and therapies to the individual. Moreover, healthcare professionals must understand the automated decision-making process to verify it. Prevention is crucial here. It maintains normal mental functioning, avoids the onset of critical conditions, prevents existing illnesses from progressing, reduces healthcare burden, and makes people satisfied and happier \cite{r5}. Reducing the gap between health and well-being is something that traditional healthcare may struggle to achieve. These technologies can help by ingraining positive health habits through sustained user engagement \cite{r6, r7}.

This paper does not focus on the implementation and performance details of the AI models and algorithms, but the consequences they have for the user experience of a specific healthcare technology employing these models. In particular, we concentrate on user-related aspects of these technologies that may influence their acceptance and the effectiveness of mental healthcare provision.

The paper is structured as follows. Section 2 describes a landscape of mental health interventions organized as robotic technologies, video games, and conversational agents that can sense and respond to human emotions. Section 3 introduces the criteria used to analyze the technologies. Section 4 presents the results of the analysis, with the future outlook. Section 5 concludes the paper.

\section{Related Work}
\label{sec:2}

We describe the intelligent interactive technologies for mental healthcare as they emerged. The systems are grouped as robotic technologies, video games, and conversational agents that can recognize and adapt to users’ emotions (being an emerging requirement for digital mental health therapy).

\subsection{Robotic technologies}
\label{subsec:1}

{\textit{Social robotics}} has proved to be a useful method for treating autism spectrum disorder (ASD) in children. Previous research attests to the various benefits of robotic devices targeting specific aspects of ASD behavior. The outcomes range from decreasing repetitive and stereotyped behaviors \cite{r8}, through enhanced attention and imitation behavior for social learning \cite{r9}, to the induction of spontaneous linguistic behavior \cite{r10,r11,r12}. The results of using socially assistive robots to improve mental health are mixed and there is a lack of empirical evidence or utility of using them in mental health interventions \cite{r13}.

One of the most popular programmable humanoid robots is NAO. It has multiple degrees of freedom and sensors, as well as the integrated speech recognition support. There is a wide range of applications of NAO robot, including interventions for children with ASD \cite{r14}, diabetes management \cite{r15}, reduction of apathy and stabilization of cognitive performance in advanced dementia patients \cite{r16}, supporting the learning of pupils with severe intellectual disabilities \cite{r17}, rehabilitation training \cite{r18}, reducing pediatric distress and pain in medical settings \cite{r19}, and logopedic interventions \cite{r20,r22}. Pepper is a humanoid robot that uses the same underlying platform. A study shows there are no differences in expectations between the models related to the physical appearance - before the interaction. However, it changes after the interaction \cite{r21}.

Pediatric care is a field in which the use of intelligent robotic technology is increasing. Huggable is a smartphone-based social robot designed to address the emotional needs of hospitalized children by engaging them in playful interactions and promoting their socio-emotional well-being \cite{r23}. It has 12 degrees of freedom and can move its head, shoulders, elbows, waist, muzzle, and ears. Another example of an Android-based robot platform that leverages smartphones to drive computation and display an animated face is Tega \cite{r24}. The platform is designed to support long-term, in-home interactions with children in early-literacy education, from vocabulary to storytelling.

Medi is another example of a social robot used in pediatrics, designed to reduce the level of anxiety and stress when children cope with pain \cite{r25}. The robot uses a set of behaviors derived from pain management literature and cognitive behavioral therapy (CBT). The built-in functions include personalized greetings, playing games, storytelling, and dancing. Robots such as the THERAPIST are used in motoric and neurorehabilitation therapies for children through different games activities activities, but require complex cognitive architectures to provide optimal and personalized experiences \cite{r26}.

CommU is a smaller humanoid robot with 14 degrees of freedom - including directed eye gaze - was used by teenagers with ASD to converse with their teacher \cite{r27} indirectly. The robot was designed to be small and cute to avoid fearfulness among the children. Regardless of the lack of ML/DL, the robot's characteristics make it a compatible for their later implementation. When designing the robot's exterior and appearance, robustness and sanitation ability are important factors for use in the hospital environment. Actroid-F is another robot used as an intermediary for communication with young adults with ASD \cite{r28}.

Toy robot Pekoppa, shaped like a bilobed plant with a responsive stem and leaves, was used as a listener in communication with neurotypical and children with autism aged 6–7 years \cite{r29}. The robot's form is compatible with the finding that children with autism show a preference for minimalist objects to which they can their own mental states or those of others \cite{r30}.

Other than the previous solutions, open-source robotic technologies are emerging. For example, the Poppy project is a community-centered robotic project that provides free specifications for 3D printing of a humanoid robot, torso, or arm \cite{r31}. The printouts can be combined with affordable controllers, including Arduino, RaspberryPi, and others. Other examples are iCub, a humanoid robot designed to support research in embodied AI \cite{r32}, and Oncilla \cite{r33}. Additional sensors could be integrated into the solution to improve the Human-Robot Interaction experience \cite{r34}.

A large user group that prefers and benefits from interacting with physically embodied robots are older people\cite{r35,r36,r37}. Social robotics developed for this user group is mainly oriented towards preventing dementia \cite{r38} and loneliness \cite{r39}. Telenoid \cite{r40} and Giraff \cite{r41} are examples of robots designed to support telepresence. The Nurse-Bot project is a RaspberryPi based robot system to provide medical assistance in later life \cite{r42}.

The main concern with robotic technologies is user acceptance, influenced by how likable and trustworthy they are. If a robot looks too human-like but does not match social expectations in terms of behavior, the user might distrust these systems \cite{r43}. Social robotics is unlikely to be ubiquitous in the near future due to its higher cost compared to other technologies, while physical presence is not a prerequisite for emotional expression \cite{r94}.

\subsection{Video games}
\label{subsec:2}

Another stream of research and system development shows the various benefits of video games for the mental health and well-being of different target user groups \cite{r44,r45,r46,r47,r48,r49,r50,r51,r52,r53,r54,r55,r56,r57}.

The most common properties of video games concerning enhanced mental well-being include feedback on progress, points and scoring, rewards and prizes, narratives, personalization, and customization \cite{r50}. The related target aspects of mental well-being are anxiety, well-being, alcohol use, and depression. On the other hand, games suffer from methodological issues concerning the validity and generalizability of their findings. Games are used as a black-box solution without understanding their underlying mechanism and how they approach and tackle specific mental health conditions. The concrete health outcomes are described in terms of perceivable user experience elements (as the ones mentioned above), and not in relation to the game mechanics, comprehensive explanations of how they are used, and causal relationships between game tasks/rules and elements of human mental functioning \cite{r50}. Resolving the latter challenges is crucial for the correct application of games in mental healthcare.

As we get older, we notice many changes. Our brain may also show signs of ageing. Cognitive decline in thinking, language, memory, understanding, and judgment are often a normal part of ageing. It can also be an early sign of dementia, a group of brain disorders characterized by a more severe, irreversible decline in cognitive functions. Engaging in activities that stimulate the brain throughout life is thought to enhance cognition in later life and reduce the risk of age-related cognitive decline and dementia.

Cognitive training is an intervention that provides structured practice on tasks relevant to different aspects of cognition, such as attention, memory, or executive functions \cite{r51}. Training can focus on a single cognitive domain (e.g., memory) or multi-domain training (including attention and processing speed). As such, video games have been used to design the cognitive training.

In cognitive/emotional training, most mental health interventions used action games (i.e., Call of Duty: Modern Warfare3 in \cite{r53}), followed by puzzle games (i.e., Angry Birds from \cite{r54}) \cite{r52}. The prevalent aspects of cognition that are trained include processing speed and reaction times, memory, task-switching/multitasking, and mental spatial rotation \cite{r52}.

One of the main difficulties in obtaining a clear picture of the effects of video game training is the significant variability of several key variables of the intervention studies - the type of video game used, the cognitive process assessed, how these cognitive processes were evaluated, and the different personal characteristics of the trainees \cite{r50, r52, r56}. This variability seems to be the leading cause of the mixed results reported in the literature \cite{r52}. The reviewed studies indicate that video game training improves some aspects of cognition but not others \cite{r50, r56}. The main limitations are relatively small sample sizes in longitudinal intervention studies, the diversity of motivational factors for playing video games, and the placebo effect resulting from familiarity with the investigators \cite{r50, r52, r56}.

Concerning specific target groups, video games are increasingly used to help autistic children improve and maintain their social and cognitive skills \cite{r45,r46,r47,r48,r49}.  For example, the ECHOES game \cite{r44} uses AI to control a virtual character during a joint attention and communication skills training program. The agent acts as a social partner to autistic children when learning. This technological intervention includes interactive storytelling elements, medium-term goals, and rewards for success. The evaluation revealed that children initiated interactions with the virtual character more often.

The children are motivated to engage with the games, find them easy to use, and learn relevant content over a short period. There are important questions about how effective and how much is learned using video games. Regarding effectiveness, games' clinical validation does not meet evidence-based medicine standards, and there is a lack of connection and compatibility between the game design and the clinical validation \cite{r48,r49}. Concerning learning, the results show that it is important provide an educational game context that meets the children's diverse interests and preferences. The majority of existing games have a positive effect on high-functioning individuals \cite{r46,r48}. In general, games represent a complex design space requiring research on how different gamification approaches can affect the experiences of participants with inherently varied motivations and interests in gaming \cite{r56}. The heterogeneity of the children with autism impacts how the user model of a serious game for different users needs to be constructed.

The success of gamified cognitive training has been demonstrated for both commercial and noncommercial video games \cite{r52,r56}. Some studies (dealing with emotion regulation) even suggest using commercial games rather than games developed with specific purposes due to prolonged engagement, which is critical for effective mental health treatment \cite{r55,r56}. On the other hand, current gamified technologies lack standardized terminology, development and study protocols, best practices in matching games and mental health tasks, and clear ethical procedures\cite{r56}.

Similarly, gamification as the practice of using game elements in non-game settings \cite{r57} has been recently proposed for digital health interventions \cite{r58}. The work described in \cite{r58} offers a model of gamification principles for digital healthcare. The principles are articulated as user requirements and cover meaningful purpose, meaningful choice, supporting player archetypes, feedback, and visibility (see \cite{r58} for details). They were empirically validated in a user study (N=113 raters) by assessing user satisfaction against Web and mobile applications (N=17). The study revealed a significant correlation between three variables (supporting player archetypes, feedback, and visibility) and the test of application quality concerning engagement, functionality, aesthetics, and information quality (known as MARS \cite{r59}).

From an AI perspective, video games are a promising technology as many features are integrated into commercial (and available) gaming environments and platforms such as Unity\footnote{Unity game engine: https://unity.com/ (Retrieved on Dec 25, 2020).}. For example, the Unity games can implement intelligent agents moving in a synthetic game environment, detecting, reacting to collisions, interacting with other agents, games' objects, and avoiding dynamic obstacles\footnote{Exploring new ways to simulate the coronavirus spread: https://bit.ly/37PQWYG (Retrieved on Dec 25, 2020).} \footnote{Unity Machine Learning Agents: https://bit.ly/34RG2zM (Retrieved on Dec 25, 2020).}. The Unity Perception uses computer vision and augmented reality to detect and interpret real-world and digital game objects and characters\footnote{Training a performant object detection ML model on synthetic data using Unity Perception tools: https://bit.ly/3hjoLVl (Retrieved on Dec 25, 2020).}. Aside from using AI algorithms, a recent achievement of games is the automated generation and annotation of training data for such algorithms at scale\footnote{Use Unity’s perception tools to generate and analyze synthetic data at scale to train your ML models: https://bit.ly/3rwofrt (Retrieved on Dec 25, 2020).}. 

\subsection{Conversational agents}
\label{subsec:3}

Technology-based health interventions are increasingly using a shared design metaphor - a personal, intelligent assistant (also known as a chatbot) that provides healthcare through natural conversation. The main reason is to make existing services more user-friendly - an agent takes a patient through a turn-taking dialog, exchanging questions and answers to complete a task, similar to how doctors do \cite{r60}. Successful examples of such agents in mental health care are Woebot \cite{r60} designed to help with depression, and Vincent, aimed at raising self-compassion \cite{r61}. The technology is integrable into popular messaging platforms, such as Facebook Messenger, Slack, Telegram, and Skype. Although current chatbots do not possess the same level of physical presence as humans, even unimodal interactions (such as text or voice) can still have behavioral significance, while being less costly to design and deploy \cite{r62,r6}. Users display positive sentiments and trust towards virtual agents that provide emotional and functional affordances and dominate the users' experience with these agents \cite{r63}. A common finding was that the agents’ functionality and social interactivity are equally important and complementary rather than separate \cite{r64}. In digital healthcare, chatbots provide low-cost, easy access to medical triage (e.g., Babylon \cite{r65} ), mental health support and well-being (e.g., Woebot \cite{r60}), and health-promoting behavior change (e.g., Florence  \cite{r66}).

Recently, conversational agents have been used successfully to automate the assessment and evaluation of psychological treatments through social support, supporting mental well-being and promoting psychological functioning \cite{r67,r61}. They act as digital avatars and communicate with their users through natural dialogue. In this way, they enable medical professionals to continue mental health counseling and therapy. Some prominent examples are Woebot \cite{r60} and Wysa \cite{r68}. Woebot \cite{r60} is a virtual mental therapist who monitors users' mood and suggests tips and mental activities. Wysa \cite{r68} offers emotional support during conversation with its users to improve their mental health and well-being. CBT is an umbrella term for various treatments that aim to enhance or re-establish normal functioning concerning specific mental states and emotions. Agents implement online, remote CBT as a guided dialogue that commences with diagnosing patients’ mental conditions. Based on the diagnosis, the agent suggests and assists in specific exercises and monitors them through structured conversations. The measures of a treatment’s progress are extracted from patients' responses as free-form text or selected options. The underlying logic analyzes and interprets patients’ input using Natural Language Processing (NLP) and ML algorithms to estimate and make successive decisions during therapy. The virtual therapist reduces the gap between health and well-being - facilitating individual resilience to mental disorders (i.e., depression, stress, and anxiety) by developing desirable self-care habits through personalized mental exercise (such as mindfulness, self-awareness, and optimism).

The requirement for emotionally sensitive chatbots assumes multimodal sensing capabilities and expressing emotions in more complex ways. Computer science methods have been applied to visual, audio, and textual data to infer emotion \cite{r69}. In many cases, this involves detecting subtle signals in high-dimensional data. While verbal and nonverbal cues both contain rich information about a person’s emotional state, researchers have found significant improvement in automated understanding of nonverbal behavior by combining signals from numerous modalities (such as speech, gestures, and language) \cite{r69}. Multimodal human behavior analysis requires a sufficient amount of high-quality, diversified datasets annotated manually for facilitating learning \cite{r70,r71,r72,r73,r74,r75}.

In general, the underlying ML/DL algorithms of mental health chatbots can learn about users’ behavior from verbal and nonverbal signals.

\textit{Verbal signals.} Linguistic patterns and word choice could be linked to a user’s affective state. The LIWC \cite{r76} software package enables the automatic extraction of linguistic style features by capturing the frequency of words from different categories. Matching a person’s linguistic style (for example, through word choice) is perhaps one of the simplest ways an assistant can be designed to emotionally bond with a person. Speech signals convey linguistic and paralinguistic information features \cite{r77, r78}. Selecting a suitable and robust set of features is a challenging task. The features that have been used in the recently published research are pitch, intensity, formants, Mel-frequency cepstral coefﬁcients (MFCCs), and ﬁlter bank energies (FBE) \cite{r78}. Tools for syntactic and semantic analysis of discourse allow for natural language understanding and generation to provide meaningful conversations with users based on context analysis \cite{r79,r80}. However, speech recognition may be difficult due to the specifics of the patients' spoken language and an insufficient corpus for training speech recognition algorithms in these specific situations \cite{r79}.

\textit{Nonverbal signals.} Facial expressions are one of the richest sources of affective information. Automated facial action coding can be performed using highly scalable frameworks \cite{r81}, allowing analysis of large datasets (for example, millions of individuals). These tools are easily accessible and relatively simple to integrate into other applications. They can even execute on resource-constrained devices enabling mobile applications of facial expression analysis. While expressed affective signals are those that are most used in social interactions, physiology plays a significant role in emotional responses. Computer systems can measure many of these signals in a way that an unaided human could not. The current state of work offers physiological emotion databases combining unimodal signals that can be used for measuring the nervous system's activities connected with emotions. They contain data including electroencephalogram (EEG), functional near-infrared (fNIR), cardiopulmonary parameters (heart and respiration rates), and skin conductance \cite{r73,r74}. Despite advances in sensing emotions, specific challenges in objective measurement remain. The sparsity and lack of specificity within unimodal cues (such as facial expression) are key reasons why multimodal affective computing systems have been found to be consistently better than unimodal ones \cite{r82}. The most common approach to label emotions is discrete categorization, which uses a set of basic emotions and the associated cognitive, physiological, and behavioral processes \cite{r83}. While several categorizations have been proposed, the most used set is the so-called “basic” list of emotions: anger, fear, sadness, disgust, surprise, and joy \cite{r83}.

\section{Analytical Framework}
\label{sec:3}

In analyzing intelligent interactive technologies for mental health, we focus on elements of Human-AI interaction. We start from and adapt Human-AI interaction guidelines to explore existing technologies \cite{r84}. The original guidelines are instrumental and prescriptive concerning principles or rules to implement in interactive AI technologies \footnote{The interactive cards with practical examples of the guidelines: https://bit.ly/3b3VMUr (Retrieved on Dec 25, 2020).} In contrast, we provide a descriptive analysis of how end-users perceive and interact with these technologies. Consequently, the dimensions are extracted, combined, and extended to capture relevant user-related aspects of mental health diagnosis, prevention, and treatment services (as summarized in Table 1).

\begin{table}[!t]
\caption{The dimensions to analyze intelligent interactive technologies for mental well-being.}
\label{tab:1}       % Give a unique label
%
% Follow this input for your own table layout
%
\begin{tabular}{p{4cm}p{7.4cm}}
\hline\noalign{\smallskip}
Dimension & Description  \\
\noalign{\smallskip}\svhline\noalign{\smallskip}
Transparency & Exposing a certain level of accessibility to the system’s data and algorithms.\\
Explainability & Clarifying what the system can do, how well the system can do it, and why the system made a decision.\\
Privacy & Protecting user data during collection, analysis and use.\\
Error management & Error prevention and recovery strategies.\\
Context awareness & Recognizing and responding to the environment.\\
Learning and personalization & Learning about user behavior and adapting and evolving accordingly.\\
Empathy and social behavior & Understanding and responding to user emotions, and exposing sociability.\\
Healthcare provision & Healthcare services for prevention, diagnosis, and treatment.\\
\noalign{\smallskip}\hline\noalign{\smallskip}
\end{tabular}
\end{table}

\paragraph{\textit{Transparency and Explainability}} %
Ground truth may not exist in making various medical decisions, including prevention, diagnosis, and treatment. The “black-box” nature of current AI medical systems (meaning that details on how their learning models work are not clear nor visible to non-technical people) can raise concerns with their users \cite{r85, r7}. One of the critical aspects of ML systems in medicine is that they do not understand context, being focused on the statistical processing of the data - they can identify phenomena but still do not explain them \cite{r86}. For example, the data can reveal that the users who received a particular treatment demonstrated improvement but cannot tell us why that happened.

There is a tradeoff between increased requirements for ML/DL accuracy and explainability - higher accuracy can be less transparent (i.e., Convolutional Neural Network, CNN), while algorithms offering clear explanations (i.e., Decision Tree classifier) can lack accurate predictions \cite{r87}.
In general, concerns regarding the lack of control in Human-AI interactions are emerging (such as decision making of the vehicle in autonomous driving). In a situation in which an AI algorithm can assist or be independent in making decisions, it will be necessary for users to understand how the algorithm came to a decision \cite{r88}. Consequently, there is an increasing demand for intelligent systems in healthcare that are not only learning about their users well (being personalized and useful) but are transparent, explainable, and understandable for medical professionals and patients \cite{r7}. These attributes are also important in building user trust and facilitating acceptance of these technologies. Currently, medical professionals use intelligent technologies as a means of support rather than a replacement. Even if future AI systems possess greater autonomy in healthcare decision making, doctors will still need to understand and follow the process.

Measuring the quality of explanations and communicating the explanations to different stakeholders (including doctors and patients) in a comprehensible and user-friendly way is an emerging requirement for medical AI systems \cite{r7, r89}. Some researchers have even proposed explanation interfaces for the underlying AI models and algorithms that can measure explainability as the quality of human-AI interaction \cite{r2}, similar to usability (describing the ease of use). These interfaces should speak using non-technical language to provide understandable and trustable explanations for their decisions to users. Often, people do not fully understand the effects of certain health behaviors which may negatively influence their motivation. Providing comprehensible explanations can facilitate understanding, raise awareness concerning benefits, and persuade users to adopt a healthy lifestyle \cite{r7}.

According to the source guidelines \cite{r84}, transparency and explainability concerning system capabilities should be stated and clarified at the start of interactions with users.

\paragraph{\textit{Privacy}} %

The issue of users’ privacy is widely acknowledged in digital healthcare \cite{r90}. The main elements include ensuring that user data is always secure, that users can control their data, and that AI systems collect only necessary information about users \cite{r91}.

ML/DL-based solutions store and use a high amount of data to provide a more advanced and more personalized service. Some systems also engage additional remote resources (usually cloud-based for data storage or processing) to improve performance. Therefore, how personal data is stored and transferred over the network is crucial - namely, what kind of encryption is used and if it is used correctly. This aspect is important through the device/solution's lifespan, including cases where it is stolen, broken, or discarded. Data anonymization functions, if applied correctly, can significantly improve privacy protection by removing personally identifiable information. If the solution uses remote resources or sends statistical data to the provider for further analysis, there is a risk of revealing the user's geographical location and daily patterns.

\paragraph{\textit{Error management}} %

Since the AI technology in medicine generally does not provide deterministic or predictable outcomes per se, managing errors is important. The occurrence of errors can be avoided by design, and systems can capture, handle, or learn from errors after they occur.

The occurrence of errors in the systems' is highlighted in the guidelines from \cite{r84}. They propose efficient support in the event of errors, dismissal and recovery, and error prevention by scoping functions based on the clear understanding of users’ goals.

\paragraph{\textit{Context awareness}} %

Humans act in a material world and spend their time in different environments (including home, work, indoors, and outdoors). Therefore, the ability to sense, interpret and respond to the user's environment is an important factor in learning about their habits and behavior \cite{r92}. Combined with AI models that learn and predict specific mental conditions, they can improve automated decision-making accuracy. The capabilities are supported by mobile and wearable technologies for continuous monitoring, diagnosis, and treatment of mental health conditions \cite{r92}.

The guidelines from \cite{r84} stress the necessity of continuous context understanding throughout Human-AI interaction. The system should provide services and information based on the user’s current task and environment in a timely manner.

\paragraph{\textit{Learning and personalization}} %

This dimension’s focus is the quality of the ML/DL algorithms' output concerning individual user’s needs. The outcome ultimately depends on the algorithm's design, including the design of the model itself, data collection, and training process. However, given the nature of our analysis, we will not analyze the technical details of these algorithms, but their effectiveness as perceived by end-users. Specifically, how well they learn about users over time (perceived as accuracy and suitability of their outputs and decisions as recommendations, instructions, or answers), and what type of control over this process they provide to their users (being customizable concerning their services to match user preferences and expectations).

The guidelines in \cite{r84} suggest a Human-AI interaction that evolves so that the system keeps a record of its users' actions, learns about the users' behaviors and adapts its services accordingly, provides timely feedback on user actions and new capabilities and is customizable in that users can communicate their preferences and control the system’s behavior.

\paragraph{\textit{Empathy and social behavior}} %

Emotions play a crucial role in our well-being and communication with other people. Vice versa, the way we communicate with others influences our feelings. Interactive intelligent computer systems that respond to social and emotional cues can be more engaging and trusted while performing complex tasks in a more socially acceptable manner \cite{r93}. The field of affective computing concerns the design and development of computer systems that sense, interpret, adapt to, and potentially respond appropriately to human emotions \cite{r69}. Some authors even propose a taxonomy of the perceivable design elements connected with particular emotions \cite{r94}. The elements can be visual, verbal, auditory, and invisible (such as tactile sensations).

The starting guidelines \cite{r84} indicate that the results should be delivered to respect the user's social context and cultural background and avoid biases and stereotypical behavior towards users.

\paragraph{\textit{Healthcare provision}} %

We analyze the type of healthcare service that the system offers as prevention, diagnosis, and treatment; the number of target conditions (single or more); and the involvement of different stakeholders (such as patients, family, friends, peers, and medical professionals) \cite{r95}.

\section{Discussion and outlook}
\label{sec:4}

Here, we analyze the state of the art technologies with respect to the dimensions from Table 1. and identify outlooks for the future.

\subsection{Discussion}
The discussion is structured according to the dimensions described in chapter 3 (Table 1).

\paragraph{\textit{Transparency and Explainability}}

Robotic technologies and video games serve two general purposes concerning support for instrumental and hedonic activities. The instrumental activities are focused on fulfilling practical goals, such as activities for daily living and specific mental therapies with robotic technologies \cite{r8,r9,r10,r11,r12,r14,r15,r16,r17,r18,r19,r20,r21,r22,r27,r28,r29,r30,r31,r32,r33,r34}. The hedonic activities are engaged in for fun and enjoyment, while the concrete health outcome is a side-effect of the gameplay \cite{r50,r51,r52,r53,r54,r55,r56,r57}. 

Concerning explainability, robotic technologies do not generally provide reasons for their decisions, instructions, or suggestions to users \cite{r14,r15,r16,r17,r18,r19,r20,r21,r22,r27,r28,r29,r30,r35,r36,r37}. The details of the underlying AI algorithms and data are not accessible to users (which may be expected since the robots have physical appearances.) \cite{r14,r15,r16,r17,r18,r19,r20,r21,r22}. 

The video games, by their nature, do not expose the 'serious' parts concerning AI mechanisms and data to their users, nor they explain the rules and decision of the gameplay \cite{r44,r45,r46,r47,r48,r49,r50,r51,r52,r53,r54,r55,r56,r57}.

On the other hand, conversational agents have dialogues with their users, and each talk may require an exchange of explanations for mutual understanding between the collocutors \cite{r60,r61,r62,r63,r64,r65,r66,r67,r68}. This understanding is critical in conversations between doctors and patients due to healthcare complexity and uncertainty. Mental chatbots, such as Wysa \cite{r68} and Woebot \cite{r60}, clarify their decisions and the reasons for asking the user for specific data to some extent \cite{r95}. The chatbots’ capabilities need to be set in advance for their users to match expectations regarding mental health provision (stating what they can do).

\paragraph{\textit{Privacy}}
Observed privacy is mainly manifested with user data collection. By analyzing the existing literature, we found that systems can collect data either explicitly or implicitly. The former assumes that users provide their data explicitly, such as data entry from the system's interface, responding to a questionnaire, or through conversations. The latter concerns using various sensors (such as cameras, eye-tracking devices, and mobile device sensors for measuring position and orientation) that monitor and record user behavior during interaction ors logging user actions for future analysis.

In robotic technologies \cite{r8,r9,r10,r11,r12,r14,r15,r16,r17,r18,r19,r20,r21,r22,r35,r36,r37,r38,r39,r40,r41} and video games \cite{r44,r45,r46,r47,r48,r49,r50,r51,r52,r53,r54,r55,r56,r57}, data collection is predominantly implicit.
Conversational agents collect data from their users explicitly during conversation sessions \cite{r60,r61,r62,r63,r64,r65,r66,r67,r68}.

Dealing with implicitly collected data requires caution as they can reveal the user's identity from biometric information \cite{r101}.

\paragraph{\textit{Error management}}

The evidence on error prevention and breakdown management is scarce in robotic technologies and video games. More specific strategies and actions concerning error prevention, support, dismissal, and recovery are not clearly described.

Dialog breakdowns are common in conversational agents as their knowledge and capabilities still do not match humans. 

A common error prevention technique is scoping the dialog as multiple questions with user choice constraints such as predefined answers \cite{r60,r68,r95}.

Error repair methods mainly involve preventing users from modifying previous inputs \cite{r60,r68,r95}. 
Although framing conversations with close-ended questions can help chatbots understand their users, user input errors can still occur.

\paragraph{\textit{Context awareness}}

Robotic technologies are meant to be used in the physical world and interact with their users in different environments (i.e., homes, schools, public spaces, healthcare facilities, etc.). The prerequisites for a successful interaction include sensing, interpreting, and responding to the user's environment. The majority of the robots are equipped with sensors that collect various data about the physical environment (i.e., light, noise, temperature, objects and structures) and the users (i.e., posture, gait, speech, facial expression and other people) \cite{r8,r9,r10,r11,r12,r14,r15,r16,r17,r18,r19,r20,r21,r22,r27,r28,r29,r30,r31,r32,r33,r34,r35,r36,r37,r38,r39,r40,r41}.

The collected information is used to supplement previously learned tasks and user behaviors to improve the quality of the robot's services. 

Video games run in a virtual environment and the majority of them do not make use of the context information \cite{r44,r45,r46,r47,r48,r49,r50,r51,r52,r53,r54,r55,r56,r57}.

Similarly, chatbots implement natural language understanding and generation solely from the underlying algorithmic data and information collected from dialogs with their users \cite{r60,r61,r62,r63,r64,r65,r66,r67,r68}.

\paragraph{\textit{Learning and personalization}}
All technologies can learn about their users and adjust mental healthcare provision for better outcomes. We highlight their specificities.

Robotic technologies learn about user activities, habits, preferences and mental health conditions. Over time, they collect data implicitly from monitoring user's activities (using multiple sensors) and recording explicit interactions with them. The collected data are continuously integrated and analyzed. The results of the analyses provide insights on the user's health conditions, based on which the system generates recommendations for activities (in case of diagnosis and prevention) or adapts instructions in ongoing training sessions (in case of therapy or rehabilitation) \cite{r8,r9,r10,r11,r12,r14,r15,r16,r17,r18,r19,r20,r21,r22,r27,r28,r29,r30,r31,r32,r33,r34,r35,r36,r37,r38,r39,r40,r41}.
Additionally, most robots provide feedback on the user's progress concerning current training and activities, and offer options for users to specify their preferences and needs from the system (enabling customization). \cite{r8,r9,r10,r11,r12,r14,r15,r16,r17,r18,r19,r20,r21,r22,r27,r28,r29,r30,r31,r32,r33,r34,r35,r36,r37,r38,r39,r40,r41}.

Video games naturally hide healthcare tasks from their users. In this sense, the customization and feedback are implemented, but mainly concerning the gameplay (i.e., fun purpose) not their healthcare role (i.e., serious purpose) \cite{r44,r45,r46,r47,r48,r49,r50,r51,r52,r53,r54,r55,r56,r57}. The games are designed so that particular gameplay tasks are matched with healthcare tasks. This way, the gameplay represents a facade for prevention, diagnosis, or therapy activities in the background. The data are collected while users are playing the game, and analyzed for the correct assessment of users' health conditions. Based on the user's leaned behaviour, the game adapts to the users' to maintain or improve the health conditions' parameters \cite{r44,r45,r46,r47,r48,r49,r50,r51,r52,r53,r54,r55,r56,r57}. Some features of the games (such as levels, activities engaging one or multiple human cognitive functions/senses/actuators, and rewards) lend themselves well to mental healthcare. However, it may not always be possible to find or design a suitable game.

Commercial gaming platforms (see section 2.2, last paragraph) are increasingly integrating AI elements into their services. We see this as a future opportunity for developing learning capabilities of digital healthcare applications.

Chatbots learn from the dialogs they conduct with their users on a regular basis (using text, speech, visual elements, or a combination of the three). They analyze user input to extract and understand intentions, create content for responses, and communicate natural language responses. \cite{r102}.
The described mechanism is used to study about user behavior to deliver mental healthcare in conversation sessions. NLP algorithms are crucial for engaging in a human-like conversation with users. In particular, correct understanding of user utterances and providing appropriate responses are critical due to the complexity and diversity of languages. The chatbots are still not capable of conversing as humans do, such as inferring meaning from the context of a complex text or speech, dealing with dialects, and recognizing metaphors, jokes, or sarcasm, etc.

\paragraph{\textit{Empathy and social behavior}}

Video games expose empathy by communicating messages containing emotional cues to their users during gameplay to incentivize their players (such as motivational tips, greetings, etc.). Overall, they are not capable of recognizing and responding to user emotions during the interactions. 

On the other hand, implementing emotionally-aware behavior in robotic technologies and conversational agents has received significant attention in research\cite{r82, r62, r93, r94, r95, r96, r97}.
A related phenomenon regarding robots is the uncanny valley effect \cite{r103} describing the relationship between the degree to which a robot resembles a human and the emotional response to such a robot. In case the robot imperfectly resembles a human, they can provoke feelings of uneasiness and repulsion in observers.

At the moment, conversational agents expose emotionally-sensitive behavior mainly from the written dialogs with their users \cite{r60, r68}. They extract sentiments from users' utterances and generate appropriate responses as they detect the change in the user's mood during the conversation. In addition, recent research suggests chatbot design elements (textual, visual, and auditory) that can elicit particular emotions from users \cite{r94}.

\paragraph{\textit{Healthcare provision}}

Robotic technologies are used in treating particular mental conditions or facilitating mental well-being as prevention. 
Some robots address several related aspects of mental functioning, such as in ASD \cite{r8, r9, r10, r11, r12, r13, r14, r27, r28}. Others tackle different conditions (i.e., NEO robot from \cite{r15, r16, r17, r18, r19, r20}).
Pediatric care robots are mainly used to preserve children's mental functioning \cite{r23, r24, r25, r26, r27, r28, r29, r30}. Robotic technologies for older adults are also applied primarily during the prevention stage of mental health \cite{r35, r36, r37, r38, r39, r40, r41, r42}.
Stakeholders other than target user groups are not directly involved the use of technologies with respect to their purpose.

Video games are mainly present in the prevention period to alleviate certain conditions (such as ASD) \cite{r44, r45, r46, r47, r48, r49} and maintain cognitive functioning in old age \cite{r50, r51, r52, r53, r54, r55, r56, r57}.
Overall, gameplay does not include stakejolders from their users' social circles (i.e., doctors, formal/informal carers, and family members).

Mental health chatbots encompass all aspects of healthcare: prevention, diagnosis, and therapy \cite{r60, r61, r62, r63, r64, r65, r66, r67, r68}. Interactions are conducted with patients. They offer sustained conversations with users in their homes. In light of the current pandemic, telemedicine is becoming more ubiquitous \cite{r104, r105} as a remote and contactless healthcare provision. Chatbots represent a fundamental part of this change' gear of that shift.

\subsection{The outlook}
\label{subsec:4}

We articulate the outlook for designing intelligent interactive technologies for mental health along three main lines.

\runinhead{Emotionally-aware systems.} The technology can be described from the user perspective, in terms of the desirable hedonic and pragmatic qualities. Hedonic or experiential attributes refer to the affective and mental affordances it should provide (such as trust, emotions, personalization, and learning from interactions). Pragmatic or instrumental qualities concern the functional affordances it should enable (such as usability or ease of use, and usefulness).

One of the main criticisms of existing technologies for mental healthcare is that they are not capable of empathy, namely recognizing users’ emotional states and tailoring responses that reflect these emotions \cite{r93, r62, r96, r97}. The lack of empathy negatively affects engagement with the technology. Sustained and regular user engagement is an prerequisite for effective health intervention and crucial for creating training data to improve algorithms for learning about users. A sucessful affective system would be one that could learn about a person’s nuanced expressions and responses, and adapt to different situations and contexts. 

These models have common requirements. Firstly, the models should be machine-readable and computable. Secondly, they should combine verbal and nonverbal cues from different modalities (such as speech, text, and vision) to improve emotion recognition accuracy.

\runinhead{Scalable and scaffolded learning about user behavior.} Cognitive psychology proposes bottom-up information processing (created from our senses) and top-down inference (based on the existing knowledge about the things from the world such as words, concepts, objects, actions, and relations) \cite{r98}.

Similarly, hybrid learning that combines top-down models (such as knowledge graphs, rule-based systems, and symbolic representations) and bottom-up approaches (such as ML and DL) can yield higher quality automatic predictions and decisions \cite{r99}, or make (healthcare) AI systems more comprehensible to their users \cite{r100}. Although algorithms use approximations of real-world data and simulate human abilities, knowledge about users can unfold over time while making the system readily understandable.
The technological framework can integrate different modalities (e.g., facial expressions, text and speech content and linguistic patterns, and physiological signals) to optimize accuracy, performance, and computational cost. 

Concerning the interpretation of emotions, different models of emotions can be employed, such as a circular, two-dimensional space in which points close to one another are highly correlated \cite{r83}. This approach may provide a continuous and fine-grained interpretation of multiple emotions at once. For example, in some cases, none of the “basic” emotion labels may apply to a specific, observed emotional response, but that response can lie somewhere within the dimensional space.
The deployed framework can generate domain-specific training data (or model) for emotion recognition applications that embrace the target user group's cognitive and socio-demographic characteristics.

\runinhead{Continuous and affordable mental healthcare provision.} A multi-stage intervention that combines in-hospital treatment and out-of-hospital care. Current healthcare systems aim for a balance among improving the health of specific populations, delivering quality care to individuals, and cost control. Computerized, personalized, and online assistance and mental health treatment can reduce the burden on healthcare providers and services while increasing their effectiveness. This approach can also support informal caregiving (at places other than hospitals, such as in the home) in mental therapy to improve its overall effectiveness. This methodology brings three significant benefits.

Such an approach can enable customized and adaptive online treatment (including a mobile version) that can combine different design elements and assess their effectiveness in improving mental health and well-being with decreased data acquisition and evaluation costs.

Subsequently, advanced evaluation and quality assurance of mental health treatments in terms of existing medical standards, user attitudes, usability, and treatment effectiveness. The possibility to engage a larger number and a broader range of participants with different levels of cognitive functioning or home bound users.

Finally, the formation of a multidisciplinary community for human-friendly AI for healthcare that brings together experts from related fields (such as medicine, artificial intelligence, human-computer interaction, and information systems). The community can raise the profile of this research within these communities.

\section{Conclusion and limitations}
\label{sec:5}

This paper provides an overview of existing interactive AI technologies for mental health and well-being. The technologies were analyzed from different aspects of Human-AI interaction. The findings can be informative for various stakeholders, including medical professionals, researchers, engineers, and end-users. In line with the existing initiatives for more inclusive and accessible healthcare \cite{r104, r105}, we consider this paper a step towards this global goal.

Being a literature/research review, the present study does not claim to be comprehensive. Instead, we summarize the existing literature and research at the time of writing and focus on the shared elements of different technologies from the Human-AI interaction perspective. We identify and describe the aspects concerning the analytical framework and derive the outlook for future research. Our analysis is based on papers and not on real systems due to their lack of availability (this excludes publicly available chatbots such as Woebot \cite{r60} and Wysa \cite{r68}). It may influence overall findings. Further analysis of the specific technologies is required.

%%%%%%%%%%%%%%%%%%%%%%%% referenc.tex %%%%%%%%%%%%%%%%%%%%%%%%%%%%%%
% sample references
% %
% Use this file as a template for your own input.
%
%%%%%%%%%%%%%%%%%%%%%%%% Springer-Verlag %%%%%%%%%%%%%%%%%%%%%%%%%%
%
% BibTeX users please use
% \bibliographystyle{}
% \bibliography{}

\begin{thebibliography}{99}%


\bibitem{r1} Du, M., Liu, N., Hu, X.: Techniques for interpretable machine learning. Communications of the ACM, 63(1), 68-77. (2019)
\bibitem{r2} Holzinger, A., Langs, G., Denk, H., Zatloukal, K., Müller, H.: Causability and explainability of artificial intelligence in medicine. Wiley Interdisciplinary Reviews: Data Mining and Knowledge Discovery, 9(4), e1312. (2019)
\bibitem{r3} Killgore, W. D., Cloonen, S. A., Taylor, E. C., Dailey, N. S.: Loneliness: A signature mental health concern in the era of COVID-19. Psychiatry Research, 113117 (2020)
\bibitem{r4} Rossi, A., Panzeri, A., Pietrabissa, G., Manzoni, G. M., Castelnuovo, G., Mannarini, S.: The anxiety-buffer hypothesis in the time of COVID-19: when self-esteem protects from the impact of loneliness and fear on anxiety and depression. Frontiers in Psychology, 11 (2020)
\bibitem{r5} Ebert, D. D., Cuijpers, P., Muñoz, R. F., Baumeister, H.: Prevention of mental health disorders using internet-and mobile-based interventions: a narrative review and recommendations for future research. Frontiers in Psychiatry, 8, 116. (2017)
\bibitem{r6} Fitzpatrick, K. K., Darcy, A., Vierhile, M.: Delivering cognitive behavior therapy to young adults with symptoms of depression and anxiety using a fully automated conversational agent (Woebot): a randomized controlled trial. JMIR mental health, 4(2), e19. (2017)
\bibitem{r7} Dragoni, M., Donadello, I., Eccher, C.: Explainable AI Meets Persuasiveness: Translating Reasoning Results Into Behavioral Change Advice. Artificial Intelligence in Medicine, 101840 (2020)
\bibitem{r8} Wainer, J., et al.: A pilot study with a novel setup for collaborative play of the humanoid robot KASPAR with children with autism. International Journal of Social Robotics, 6(1), pp. 45-65. (2014)
\bibitem{r9} Zheng, Z., et al.: Robot-mediated imitation skill training for children with autism. IEEE Transactions on Neural Systems and Rehabilitation Engineering, 24(6), pp. 682-691. (2015)
\bibitem{r10} Kim, E. S., et al.: Social robots as embedded reinforcers of social behavior in children with autism. Journal of Autism and Developmental Disorders, 43(5), pp. 1038-1049. (2013)
\bibitem{r11} DE-Enigma: Playfully empowering autistic children. EC-funded research project. \\ \url{http://de-enigma.eu/. Cited 25 Dec 2020}: .
\bibitem{r12} DREAM: Development of Robot-Enhanced therapy for children with AutisM spectrum disorders. EC-funded research project. \\ \url{rhttps://www.dream2020.eu. Cited 25 Dec 2020}: .
\bibitem{r13} Scoglio, A. A., Reilly, E. D., Gorman, J. A., Drebing, C. E.: Use of social robots in mental health and well-being research: Systematic review. Journal of Medical Internet Research, 21(7), e13322. (2019) https://doi.org/10.2196/13322
\bibitem{r14} Desideri, L., Negrini, M., Malavasi, M., Tanzini, D., Rouame, A., Cutrone, M. C., Bonifacci, P., Hoogerwerf, E.-J.: Using a humanoid robot as a complement to interventions for children with autism spectrum disorder: A pilot study. Advances in Neurodevelopmental Disorders, 2(3), 273–285. (2018) https://doi.org/10.1007/s41252-018-0066-4
\bibitem{r15} Robinson N.L., Connolly J., Hides L., Kavanagh D.J.: A Social Robot to Deliver an 8-Week Intervention for Diabetes Management: Initial Test of Feasibility in a Hospital Clinic. In: Wagner A.R. et al. (eds) Social Robotics. ICSR 2020. Lecture Notes in Computer Science, vol 12483. Springer, Cham. (2020) https://doi.org/10.1007/978-3-030-62056-1\_52
\bibitem{r16} Sánchez, S. M., Mora-Simon, S., Herrera-Santos, J., Roncero, A. O., Corchado, J. M.: Intelligent Dolls and robots for the treatment of elderly people with dementia. ADCAIJ: Advances in Distributed Computing and Artificial Intelligence Journal, 9(1), 99-112. (2020)
\bibitem{r17} Aslam, S., Standen, P. J., Shopland, N., Burton, A., Brown, D.: A comparison of humanoid and non-humanoid robots in supporting the learning of pupils with severe intellectual disabilities. (2016)
\bibitem{r18} Assad-Uz-Zaman, M., Rasedul Islam, M., Miah, S., Rahman, M. H.: NAO robot for cooperative rehabilitation training. Journal of Rehabilitation and Assistive Technologies Engineering, 6, 2055668319862151. (2019)
\bibitem{r19} Trost, M. J., Ford, A. R., Kysh, L., Gold, J. I., Matarić, M.: Socially assistive robots for helping pediatric distress and pain. The Clinical Journal of Pain, 35(5), 451-458. (2019)
\bibitem{r20} Egido-García, V., Estévez, D., Corrales-Paredes, A., Terrón-López, M.-J., Velasco-Quintana, P.-J.: Integration of a social robot in a pedagogical and logopedic intervention with children: A case study. Sensors, 20(22), 6483. (2020) https://doi.org/10.3390/s20226483
\bibitem{r22} Polycarpou, Panayiota Andreeva, Anna Ioannou, Andri Zaphiris, Panayiotis.: Don’t Read My Lips: Assessing Listening and Speaking Skills Through Play with a Humanoid Robot. HCI International 2016 -- Posters' Extended Abstracts, 618. 255-260. (2016) https://10.1007/978-3-319-40542-1\_41
\bibitem{r21} Manzi, F., Massaro, D., Di Lernia, D., Maggioni, M. A., Riva, G., Marchetti, A.: Robots Are Not All the Same: Young Adults' Expectations, Attitudes, and Mental Attribution to Two Humanoid Social Robots. Cyberpsychology, Behavior, and Social Networking. (2020)
\bibitem{r23} Jeong, S., Breazeal, C., Logan, D., Weinstock, P.: Huggable: The impact of embodiment on promoting socio-emotional interactions for young pediatric inpatients. Proceedings of the 2018 CHI Conference on Human Factors in Computing Systems, 1–13. (2018) https://doi.org/10.1145/3173574.3174069
\bibitem{r24} Westlund, J. K. et al.,: Tega: A social robot. 11th ACM/IEEE International Conference on Human-Robot Interaction (HRI), 561–561. (2016) https://doi.org/10.1109/HRI.2016.7451856
\bibitem{r25} Aghel Manesh, S., Beran, T., Sharlin, E., Greenberg, S.: Medi, human robot interaction in pediatric health. CHI ’14 Extended Abstracts on Human Factors in Computing Systems, 153–154. (2014) https://doi.org/10.1145/2559206.2579529
\bibitem{r26} THERAPIST: Towards an Autonomous Socially Interactive Robot for Motor and Neurorehabilitation Therapies for Children. JMIR Rehabil Assist Technol. (2014)
\bibitem{r27} Shimaya, J., Yoshikawa, Y., Matsumoto, Y., Kumazaki, H., Ishiguro, H., Mimura, M., Miyao, M.: Advantages of indirect conversation via a desktop humanoid robot: Case study on daily life guidance for adolescents with autism spectrum disorders. (2016) https://doi:10.1109/roman.2016.7745215
\bibitem{r28} Kumazaki, H., Warren, Z., Corbett, B. A., Yoshikawa, Y., Matsumoto, Y., Higashida, H., Kikuchi, M.: Android robot-mediated mock job interview sessions for young adults with autism spectrum disorder: a pilot study. Frontiers in psychiatry, 8, 169. (2017)
\bibitem{r29} Giannopulu, I., Montreynaud, V., Watanabe, T.: Minimalistic toy robot to analyze a scenery of speaker–listener condition in autism. Cognitive Processing, 17(2), 195–203. (2016) https://doi.org/10.1007/s10339-016-0752-y
\bibitem{r30} Giannopulu, I., Montreynaud, V., Watanabe, T.: Neurotypical and autistic children aged 6 to 7 years in a speaker–listener situation with a human or a minimalist InterActor robot. In: Proceedings of IEEE RO-MAN, pp 942–947 (2014)
\bibitem{r31} Poppy Project - Open source robotic platform. \\ \url{https://www.poppy-project.org/. Cited 25 Dec 2020}: .
\bibitem{r32} Metta, G., Sandini, G., Vernon, D., Natale, L., Nori, F.: The iCub humanoid robot: an open platform for research in embodied cognition. (2008)
\bibitem{r33} Sproewitz, A., Kuechler, L., Tuleu, A., Ajallooeian, M., D'Haene, M., Moeckel, R., Ijspeert, A.: Oncilla robot: a light-weight bio-inspired quadruped robot for fast locomotion in rough terrain. (2011)
\bibitem{r34} Zhao, N., Zhang, Z., Wang, Y., Wang, J., Li, B., et al.: See your mental state from your walk: Recognizing anxiety and depression through Kinect-recorded gait data. (2019) https://doi.org/10.1371/journal.pone.0216591
\bibitem{r35} Fan, J., Beuscher, L., Newhouse, P. A., Mion, L. C., Sarkar, N.: A robotic coach architecture for multi-user human-robot interaction (RAMU) with the elderly and cognitively impaired. (2016) https://doi:10.1109/roman.2016.7745157
\bibitem{r36} Tapus, A.: The role of the physical embodiment of a music therapist robot for individuals with cognitive impairments: Longitudinal study. Virtual Rehabilitation International Conference, Haifa (2009)
\bibitem{r37} Cavallo, F., Esposito, R., Limosani, R., Manzi, A., Bevilacqua, R., Felici, E., Di Nuovo, A., Cangelosi, A., Lattanzio, F., Dario, P.: Robotic Services Acceptance in Smart Environments With Older Adults: User Satisfaction and Acceptability Study. J Med Internet Res, 20(9):e264: (2018)
\bibitem{r38} Kang, H. S., Makimoto, K., Konno, R., Koh, I. S.: Review of outcome measures in PARO robot intervention studies for dementia care. Geriatric Nursing, 41(3), 207–214. (2020) https://doi.org/10.1016/j.gerinurse.2019.09.003
\bibitem{r39} Low, J. T. S.: LECABot, mini robot companion for elderly. (2020)
\bibitem{r40} Kuwamura, K., Nishio, S., Sato, S.: Can we talk through a robot as if face-to-face? Long-term fieldwork using teleoperated robot for seniors with Alzheimer's disease. Frontiers in psychology, 7, 1066. (2016)
\bibitem{r41} Orlandini, A., Kri`ffersson, A., Almquist, L., Björkman, P., Cesta, A., Cortellessa, G., ... Loutfi, A.: Excite project: A review of forty-two months of robotic telepresence technology evolution. Presence: Teleoperators and Virtual Environments, 25(3), 204-221. (2016)
\bibitem{r42} Alvarez, J., Campos, G., Enríquez, V., Miranda, A., Rodriguez, F., Ponce, H.: Nurse-bot: a robot system applied to medical assistance. In 2018 International Conference on Mechatronics, Electronics and Automotive Engineering (ICMEAE), pp. 56-59 IEEE. (2018)
\bibitem{r43} Hortensius, R., Hekele, F., Cross, E. S.: The perception of emotion in artificial agents. IEEE Transactions on Cognitive and Developmental Systems, 10(4), pp. 852-864. (2018)
\bibitem{r44} Porayska-Pomsta, K., et al.: Blending human and artificial intelligence to support autistic children’s social communication skills. ACM Transactions on Computer-Human Interaction (TOCHI), 25(6), 35. (2018)
\bibitem{r45} Tsikinas, S., Xinogalos, S.: Studying the effects of computer serious games on people with intellectual disabilities or autism spectrum disorder: A systematic literature review. Journal of Computer Assisted Learning, 35(1), pp. 61-73. (2019)
\bibitem{r46} Malinverni, L., et al.: An inclusive design approach for developing video games for children with autism spectrum disorder. Computers in Human Behavior, 71, pp. 535-549. (2017)
\bibitem{r47} Ip, H. H., et al.: Enhance emotional and social adaptation skills for children with autism spectrum disorder: A virtual reality enabled approach. Computers \& Education, 117, pp. 1-15. (2018)
\bibitem{r48} Grossard, C., Grynspan, O., Serret, S., Jouen, A. L., Bailly, K., Cohen, D.: Serious games to teach social interactions and emotions to individuals with autism spectrum disorders (ASD Computers \& Education, 113, pp. (2017)
\bibitem{r49} Chen, J., et al.: A pilot study on evaluating children with autism spectrum disorder using computer games. Computers in Human Behavior, 90, pp. 204-214. (2019)
\bibitem{r50} Cheng, V. W. S., Davenport, T., Johnson, D., Vella, K., Hickie, I. B.: Gamification in apps and technologies for improving mental health and well-being: systematic review. JMIR mental health, 6(6), e13717. (2019)
\bibitem{r51} Ballesteros, S., Kraft, E., Santana, S., Tziraki, C.: Maintaining older brain functionality: a targeted review. Neuroscience \& Biobehavioral Reviews, 55, 453-477. (2015)
\bibitem{r52} Pallavicini, F., Ferrari, A., Mantovani, F.: Video games for well-being: A systematic review on the application of computer games for cognitive and emotional training in the adult population. Frontiers in psychology, 9, 2127. (2018)
\bibitem{r53} Hutchinson, C. V., Barrett, D. J., Nitka, A., Raynes, K.: Action video game training reduces the Simon Effect. Psychonomic bulletin \& review, 23(2), 587-592. (2016)
\bibitem{r54} Clemenson, G. D., Stark, C. E.: Virtual environmental enrichment through video games improves hippocampal-associated memory. Journal of Neuroscience, 35(49), 16116-16125. (2015)
\bibitem{r55} Villani, D., Carissoli, C., Triberti, S., Marchetti, A., Gilli, G., Riva, G.: Video games for emotion regulation: a systematic review. Games for health journal, 7(2), 85-99. (2018)
\bibitem{r56} Colder Carras, M., Van Rooij, A. J., Spruijt-Metz, D., Kvedar, J., Griffiths, M. D., Carabas, Y., Labrique, A.: Commercial video games as therapy: A new research agenda to unlock the potential of a global pastime. Frontiers in psychiatry, 8, 300. (2018)
\bibitem{r57} Deterding, S., Dixon, D., Khaled, R., Nacke, L.: From game design elements to gamefulness: defining" gamification". In Proceedings of the 15th international academic MindTrek conference: Envisioning future media environments, pp. 9-15. (2011)
\bibitem{r58} Floryan, M., Chow, P. I., Schueller, S. M., Ritterband, L. M.: The Model of Gamification Principles for Digital Health Interventions: Evaluation of Validity and Potential Utility. Journal of Medical Internet Research, 22(6), e16506. (2020)
\bibitem{r59} Stoyanov, S. R., Hides, L., Kavanagh, D. J., Zelenko, O., Tjondronegoro, D., Mani, M.: Mobile app rating scale: a new tool for assessing the quality of health mobile apps. JMIR mHealth and uHealth, 3(1), e27. (2015)
\bibitem{r60} Woebot - a text-based, virtual companion who advises people about their mental well-being. \\ \url{https://woebot.io/. Cited 25 Dec 2020}: .
\bibitem{r61} Lee, M., et al.: Caring for Vincent: A Chatbot for Self-Compassion. In Proceedings of the 2019 CHI Conference on Human Factors in Computing System, 702, ACM. (2019)
\bibitem{r62} Yang, X., Aurisicchio, M., Baxter, W.: Understanding Affective Experiences With Conversational Agents. In Proceedings of the 2019 CHI Conference on Human Factors in Computing Systems, p. 542, ACM. (2019)
\bibitem{r63} Zhou, M. X., et al.: Trusting Virtual Agents: The Effect of Personality. ACM Transactions on Interactive Intelligent Systems (TiiS), 9(2-3), p. 10. (2019)
\bibitem{r64} Jain, M., et al.: Evaluating and informing the design of chatbots. In Proceedings of the 2018 Designing Interactive Systems Conference, pp. 895-906, ACM. (2018)
\bibitem{r65} Babylon - accessible and affordable health service. \\ \url{https://www.babylonhealth.com/. Cited 25 Dec 2020}: .
\bibitem{r66} Florence - personal health assistant. \\ \url{https://florence.chat/. Cited 25 Dec 2020}: .
\bibitem{r67} Miner, A., et al.: Conversational agents and mental health: Theory-informed assessment of language and affect. In Proceedings of the 4th ACM Intern. Conf. Human Agent Interaction, pp. 123–130. (2016)
\bibitem{r68} Inkster, B., Sarda, S., Subramanian, V.: An empathy-driven, conversational artificial intelligence agent (Wysa) for digital mental well-being: real-world data evaluation mixed-methods study. JMIR mHealth and uHealth, 6(11), e12106. (2018)
\bibitem{r69} Poria, S., et al.: A review of affective computing: From unimodal analysis to multimodal fusion. Information Fusion, 37, pp. 98-125. (2017)
\bibitem{r70} Jackson, P., Haq, S.: Surrey audio-visual expressed emotion (SAVEE) database. University of Surrey: Guildford, UK. (2014)
\bibitem{r71} Martin, O., Kotsia, I., Macq, B., Pitas, I.: The eNTERFACE'05 audio-visual emotion database. In 22nd International Conference on Data Engineering Workshops (ICDEW'06), pp. 8-8. IEEE. (2006)
\bibitem{r72} Wang, Y., Guan, L.: Recognizing human emotional state from audiovisual signals. IEEE Transactions on Multimedia, 10(5), pp. 936-946. (2008)
\bibitem{r73} Correa, J. A. M., et al.: Amigos: a dataset for affect, personality and mood research on individuals and groups. (2018)
\bibitem{r74} Song, T., et al.: MPED: A multi-modal physiological emotion database for discrete emotion recognition. IEEE Access, 7, pp. 12177-12191. (2019)
\bibitem{r75} Ong, D., Wu, Z., Tan, Z. X., Reddan, M., Kahhale, I., Mattek, A., Zaki, J.: Modeling emotion in complex stories: the Stanford Emotional Narratives Dataset. (2019)
\bibitem{r76} LIWC (Linguistic Inquiry and Word Count) \\ \url{http://liwc.wpengine.com/ Retrieved on Dec 25, 2020.}: .
\bibitem{r77} Schuller, B., et al.: Paralinguistics in speech and language-State-of-the-art and the challenge. Computer Speech \& Language, 27(1), pp. 4-39. (2013)
\bibitem{r78} Kamińska, D., Sapiński, T., Anbarjafari, G.: Efficiency of chosen speech descriptors in relation to emotion recognition. EURASIP Journal on Audio, Speech, and Music Processing, 3 (2017)
\bibitem{r79} Hwang, I., Lee, Y., Yoo, C., Min, C., Yim, D., Kim, J.: Towards Interpersonal Assistants: Next-Generation Conversational Agents. IEEE Pervasive Computing, 18(2), pp. 21-31. (2019)
\bibitem{r80} Abdul-Kader, S. A., Woods, J. C.: Survey on chatbot design techniques in speech conversation systems. International Journal of Advanced Computer Science and Applications, 6(7), pp. 72-80. (2015)
\bibitem{r81} McDuff, D., Girard, J. M., El Kaliouby, R.: Large-scale observational evidence of cross-cultural differences in facial behavior. Journal of Nonverbal Behavior, 41(1), pp. 1-19. (2017)
\bibitem{r82} McDuff, D., et al.: A Multimodal Emotion Sensing Platform for Building Emotion-Aware Applications. (2019)
\bibitem{r83} Marsella, S., Gratch, J., Petta, P.: Computational models of emotion. Blueprint for Affective Computing - A Sourcebook and Manual. Oxford University Press, pp. 21-46. (2010)
\bibitem{r84} Amershi, S., et al.,: Guidelines for Human-AI Interaction. (2019)
\bibitem{r85} The IEEE Global Initiative on Ethics of Autonomous and Intelligent Systems,: Ethically aligned design: A vision for prioritizing human well-being with autonomous and intelligent systems. Affective Computing, pp. 90-109. First Edition. IEEE. (2019) Retrieved on Dec 25, 2020 from https://bit.ly/3pTNtzv
\bibitem{r86} Pearl, J., Mackenzie, D.: The book of why: the new science of cause and effect. (2018)
\bibitem{r87} Bologna, G., Hayashi, Y.: Characterization of symbolic rules embedded in deep DIMLP networks: a challenge to transparency of deep learning. Journal of Artificial Intelligence and Soft Computing Research, 7(4), 265-286. (2017)
\bibitem{r88} Adadi, A., Berrada, M.: Peeking inside the black-box: A survey on Explainable Artificial Intelligence (XAI IEEE Access, 6, 52138-52160. (2018)
\bibitem{r89} Miller, T.: Explanation in artificial intelligence: Insights from the social sciences. Artificial Intelligence, 267, 1-38. (2019)
\bibitem{r90} Nurgalieva, L., O’Callaghan, D., Doherty, G.: Security and Privacy of mHealth Applications: A Scoping Review. IEEE Access, 8, 104247-104268. (2020)
\bibitem{r91} Broadbent, E.: Interactions with robots: The truths we reveal about ourselves. Annual review of psychology, 68, 627-652. (2017)
\bibitem{r92} Bardram, J. E., Matic, A.: A decade of ubiquitous computing research in mental health. IEEE Pervasive Computing, 19(1), 62-72. (2020)
\bibitem{r93} McDuff, D., Czerwinski, M.: Designing emotionally sentient agents. Communications of the ACM, 61(12), pp. 74-83. (2018)
\bibitem{r94} Feine, J., Gnewuch, U., Morana, S., Maedche, A.: A taxonomy of social cues for conversational agents. International Journal of Human-Computer Studies, 132, 138-161. (2019)
\bibitem{r95} Jovanovic, M., Baez, M., Casati, F.: Chatbots as conversational healthcare services. IEEE Internet Computing. (2020)
\bibitem{r96} Nadarzynski, T., Miles, O., Cowie, A., Ridge, D.: Acceptability of artificial intelligence (AI)-led chatbot services in healthcare: A mixed-methods study. Digital Health, 5 (2019)
\bibitem{r97} Grudin, J., Jacques, R.: Chatbots, Humbots, and the Quest for Artificial General Intelligence. In Proceedings of the 2019 CHI Conference on Human Factors in Computing Systems, p. 209, ACM. (2019)
\bibitem{r98} Delorme, A., Rousselet, G. A., Macé, M. J. M., Fabre-Thorpe, M.: Interaction of top-down and bottom-up processing in the fast visual analysis of natural scenes. Cognitive Brain Research, 19(2), 103-113. (2004)
\bibitem{r99} Mao, J., Gan, C., Kohli, P., Tenenbaum, J. B., Wu, J.: The neuro-symbolic concept learner: Interpreting scenes, words, and sentences from natural supervision. (2019)
\bibitem{r100} Gaur, M., Faldu, K., Sheth, A.: Semantics of the Black-Box: Can knowledge graphs help make deep learning systems more interpretable and explainable?. (2020)
\bibitem{r101} Neal, T., Sundararajan, K., Woodard, D.: Exploiting linguistic style as a cognitive biometric for continuous verification. 2018 International Conference on Biometrics (ICB), 270–276. (2018) https://doi.org/10.1109/ICB2018.2018.00048
\bibitem{r102} Harms, J. G., Kucherbaev, P., Bozzon, A., Houben, G. J.: Approaches for dialog management in conversational agents. IEEE Internet Computing, 23(2), 13-22. (2019)
\bibitem{r103} Mori, M., MacDorman, K. F., Kageki, N.: The uncanny valley [from the field]. IEEE Robotics Automation Magazine, 19(2), 98–100. (2012) https://doi.org/10.1109/MRA.2012.2192811
\bibitem{r104} Khemapech, Ittipong.: Telemedicine – Meaning, Challenges and Opportunities. (2019)
\bibitem{r105} Craig, J., Patterson, V.: Introduction to the practice of telemedicine. Journal of Telemedicine and Telecare, 11(1), 3–9. (2005) https://doi.org/10.1258/1357633053430494
\end{thebibliography}
%

\end{document}